\documentclass[10pt,journal,letterpaper,compsoc]{IEEEtran}

\newcommand{\eg}{{\it e.g., }}

\newcommand{\name}{{E2C }}

\usepackage{balance}
\usepackage{times}
\usepackage{cite} 
\usepackage{pifont}
\usepackage{color}
\usepackage{epsfig}
\usepackage{float}
\usepackage{graphicx} 
\usepackage{amsmath}
\usepackage{url}
\usepackage{subfigure}
\usepackage{amsthm}
\usepackage{algorithmicx}
\usepackage{algpseudocode}
\usepackage{algorithm}
\usepackage{xspace}
\usepackage{comment}
\usepackage[colorlinks, citecolor=blue]{hyperref}

\begin{document}
\title{E2C: A Visual Simulator for Heterogeneous Computing Systems}

\author{Ali Mokhtari, Mohsen Amini Salehi\\

High Performance Cloud Computing (HPCC) Laboratory,\\ School of Computing and Informatics,\\ University of Louisiana at Lafayette, LA 70503, USA\\
 E-mail: \{ali.mokhtari1, amini\}@louisiana.edu
    }

\maketitle
\begin{abstract}
 Heterogeneity has been an indispensable aspect of distributed computing throughout the history of these systems. In particular, with the increasing prevalence of accelerator technologies (e.g., GPUs and TPUs) and the emergence of domain-specific computing via ASICs and FPGA, the matter of heterogeneity and harnessing it has become a more critical challenge than ever before. Harnessing system heterogeneity has been a longstanding challenge in distributed systems and has been investigated extensively in the past. Making use of real infrastructure (such as those offered by the public cloud providers) for benchmarking the performance of heterogeneous machines, for different applications, with respect to different objectives, and under various workload intensities is cost- and time-prohibitive. To mitigate this burden, we develop an open-source simulation tool, called E2C, that can help researchers and practitioners study any type of heterogeneous (or homogeneous) computing system and measure its performance under various system configurations. E2C has an intuitive graphical user interface (GUI) that enables its users to easily examine system-level solutions (scheduling, load balancing, scalability, etc.) in a controlled environment within a short time and at no cost. In particular, E2C offers the following features: (i) simulating a heterogeneous computing system; (ii) implementing a newly developed scheduling method and plugging it into the system, (iii) measuring energy consumption and other output-related metrics; and (iv) powerful visual aspects to ease the learning curve for students. Potential users of E2C can be undergraduate and graduate students in computer science/engineering, researchers, and practitioners.
\end{abstract}
\section{Introduction}
Heterogeneity has been an indispensable aspect of distributed computing throughout the history of these systems. In the modern era, as Moore's law is losing momentum due to the power density and heat dissipation limitations~\cite{taylor2012dark, esmaeilzadeh2011dark}, heterogeneous computing systems have attracted even more attention to overcome the slowdown in Moore's law and fulfilling the desire for higher performance in various types of distributed systems. In particular, with the increasing prevalence of accelerator technologies (\eg GPUs and TPUs) and the emergence of domain-specific computing via ASICs \cite{taylor2020asic} and FPGA \cite{bobda2022future}, the matter of heterogeneity and harnessing it has become a more critical challenge than ever before to deal with. 

Examples of heterogeneity can be found in any type of distributed system. Public cloud providers offer and operate based on a wide variety of machine types. Hyperscalers such as AWS and Microsoft Azure provide computing services ranging from general-purpose X86-based and ARM-based machines to FPGAs and accelerators~\cite{Amazon_SageMaker}. In the context of Edge computing, domain-specific accelerators (ASICs and FPGA) and general-purpose processors are commonly used together to perform near-data processing, thereby unlocking various real-time use cases (\eg edge AI and AR/VR applications)~\cite{google_glass, Qualcomm}. In the HPC context, deploying various machine types with different architectures on HPC boards to fulfill the power and performance requirements is becoming a trend~\cite{cardwell2020truly}.

Heterogeneity plays a key role in improving various performance objectives of distributed systems, such as cost, energy consumption, and QoS. That is why harnessing system heterogeneity has been a longstanding challenge in distributed systems and has been investigated extensively in the past (\eg \cite{mokhtari2020autonomous,DENNINNART202046,salehi2016stochastic,liperformanceanalysis,li2018cost,ipdps19}). Making use of real infrastructure (such as those offered by the public cloud providers) for benchmarking the performance of heterogeneous machines, for different applications, with respect to different objectives, and under various workload intensities is cost- and time-prohibitive. As an example, consider an IoT-based system that offers multiple smart applications to its users (\eg object detection, face recognition, speech recognition, etc.); there exists a wide range of machine types with different architectures (such as x-86 or ARM-based multi-core CPUs, different types of GPUs, FPGAs, and ASICs) that can process these services. To find an optimal configuration, ideally, all permutations of these configurations must be examined. Moreover, there can be multiple workload intensities and scheduling policies that can affect the performance of the system. In addition, the energy consumption of the system might be of interest, which in turn, it adds another dimension to the evaluation tests. 

To mitigate the burden of examining all cases, we need simulation tools that can help us study the performance of various system configurations. To that end, in this tutorial, we introduce \name that is a framework to simulate any type of heterogeneous computing system. By using \name, the researchers can easily examine their system-level solutions (scheduling, load balancing, scalability, etc.) in a controlled environment within a short time and at no cost. In particular, \name offers the following features: (i) simulating a heterogeneous computing system; (ii) implementing a newly developed scheduling method and plugging it into the system, (iii) measuring power and other output-related things, and (iv) visual aspects to ease the learning curve for students.
These features help researchers who study resource allocation solutions in distributed systems to test and evaluate their solutions easier and faster. Moreover, the graphical user interface would help students to gain a deeper knowledge of resource allocation procedures in distributed computing systems. 
In the following sections, we elaborate on these features in more detail and discuss why UCC is a good fit for this work.

\section{Simulating a Heterogeneous Computing System via \name}
\begin{figure*}
    \centering
    \includegraphics[ height=0.3\textwidth]{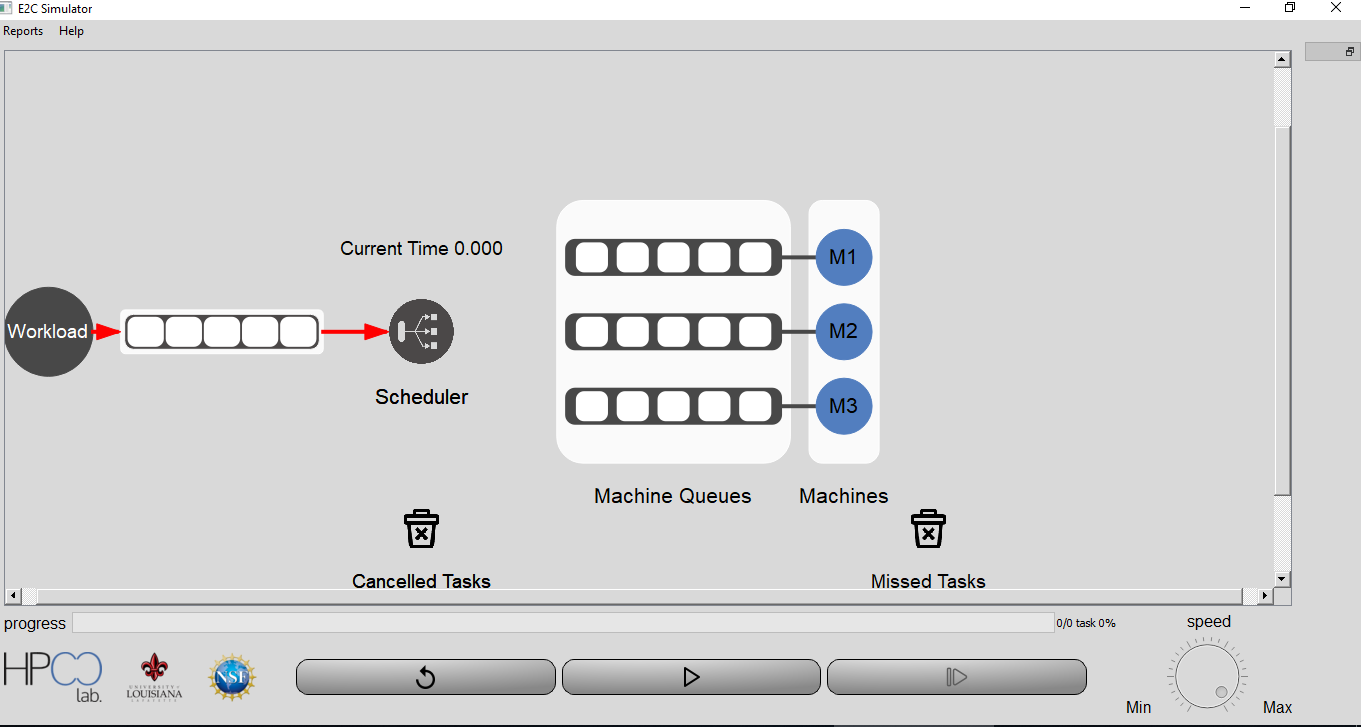}
    \caption{Overview of the \name Simulator}
    \label{fig:overviewl}
\end{figure*}

Figure\ref{fig:overviewl} shows the overview of the \name. The major components in \name are: (i) workload, (ii) batch queue (iii) scheduler, (iv) machine queue, and (v) machines. In addition, there are two more components that contain canceled and dropped tasks.

\begin{figure}
    \centering
    \includegraphics[width=0.45\textwidth]{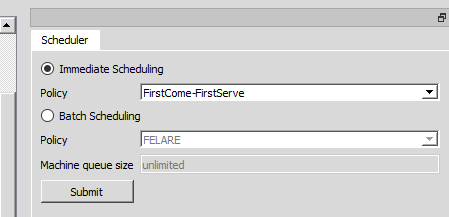}
    \caption{Workload component. The user can modify the EET matrix and arrival times of the task via the workload component.}
    \label{fig:workload}
\end{figure}

Note that the heterogeneity of the system is modeled by the Expected Execution Time (EET) matrix. As shown in Figure~\ref{fig:workload}, the user has access to EET matrix by selecting the workload component. Users can either modify the EET matrix manually or load the desired one as a CSV file. 

The arrival of a task is also generated in the workload component. As shown in Figure~\ref{fig:workload}, the user can load the desired workload trace in this section.

\begin{figure}
    \centering
    \includegraphics[width=0.45\textwidth]{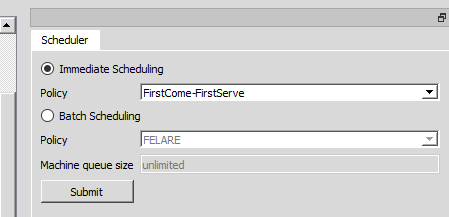}
    \caption{The scheduler options. The user can change the scheduling method here. The user should also set the machine queue size for batch scheduling methods.}
    \label{fig:scheduler}
\end{figure}

Upon the arrival of a task, the simulator transfers the task to the batch queue. Next, based on the selected scheduling method, the scheduler will select a task from the batch queue. Figure~\ref{fig:scheduler} shows the scheduler options.

\begin{figure}
    \centering
    \includegraphics[width=0.45\textwidth]{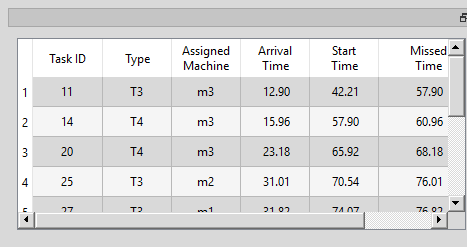}
    \caption{Missed Tasks component shows tasks that missed their deadline. }
    \label{fig:missed}
\end{figure}

The user can modify the scheduling method by selecting the scheduler component. There exist two options for the selected tasks: (i) it might be canceled or (ii) it might be mapped to one of the available machines. The status of a canceled task is set to ``canceled'' and no more process is needed. The canceled tasks component shows the tasks have been canceled so far. In the case of mapping decisions, the task is appended to the local queue of the assigned machine. Tasks are executed on the assigned machine in a sequential manner by default. If a task missed its deadline while executing on the machine, it is dropped from the machine. As shown in Figure~\ref{fig:missed}, the Missed Tasks component shows the tasks that missed their deadline. 

The simulator can be used for instructional and research purposes. So far, we have used the E2C simulator for undergraduate and graduate Distributed Computing courses to examine the impact of different scheduling policies on homogeneous and heterogeneous systems with various workload intensities. Similarly, the simulator can be used for the operating system course at the undergraduate and graduate levels to teach students about scheduling methods. The simulator has also been used to implement two research works. In \cite{ali22}, we have used E2C to examine energy efficiency and fairness of scheduling methods on a heterogeneous edge. In \cite{zobaed22}, we extended E2C to simulate the memory allocation policies of multi-tenant applications on a homogeneous edge. 

\section{Potential Users of E2C Simulator}
Students learning distributed systems can benefit from \name simulator. They can deeply investigate scheduling methods and gain insights into their advantages and disadvantages. They can also learn how heterogeneity can improve the performance of the system. Moreover, they can study the energy consumption of the system once a certain scheduling method is applied. Researchers in resource allocation area and cloud solution architects can employ \name simulator to test their solution prior to implementation. 

\section{Biography of the instructor(s)}
Dr. Mohsen Amini Salehi is an Associate Professor and Francis Patrick Clark/BORSF Endowed Professorship holder at the School of Computing and Informatics (CMIX), University of Louisiana Lafayette, USA. He is the director of High Performance and Cloud Computing (HPCC) Laboratory where several graduate and undergraduate students research on various aspects of Distributed and Cloud computing. Dr. Amini is an NSF CAREER Awardee and, so far, he has had 11 research projects funded by National Science Foundation (NSF) and Board of Regents of Louisiana. He has also received five awards and certificates from University of Louisiana at Lafayette in recognition of his innovative research.  Dr. Amini has been an active researcher in Distributed and Cloud computing research areas since 2004. He has received several awards in recognition of his research, including the “Best Intern Award” from Infosys Ltd., in 2012, and the IEEE award for “Applied Research”, in 2009. His research interests are in building smart systems across edge-to-cloud continuum, virtualization, resource allocation, energy-efficiency, heterogeneity, and trustworthy in Distributed, Edge, and Cloud computing systems.

Ali Mokhtari is currently a Ph.D. candidate in computer science at the University of Louisiana at Lafayette. Ali works as a research assistant at HighPerformance Cloud Computing (HPCC) Lab in the computer science
department. His research interest includes using reinforcement learning for
efficient resource allocation in heterogeneous edge computing systems. He
earned an M.Sc. in aerospace engineering from the Sharif University of
Technology in Iran and a B.Sc. in mechanical engineering from Shiraz
University in Iran.

\section{\name Code and Resource Availability}
\name core is available for download at the following address:
\smallskip\noindent
\footnotesize \texttt{\url{https://github.com/hpcclab/E2C-Sim}}
\normalsize

\smallskip \noindent
The manual document on how to run \name and its options and full documentations are available here:

\smallskip\noindent
\footnotesize \texttt{\url{https://hpcclab.github.io/E2C-Sim-docs/}}
\normalsize

\smallskip \noindent
The video resources for \name are in this  \href{https://youtube.com/playlist?list=PL7jhdCPVrCHh49PvIglDEY2Xs4v2ivrsw}{YouTube page}.

\section{acknowledgement}
Development of E2C was made possible by the funding support provided by National Science Foundation (NSF) under awards\# CNS-2007209 and CNS-2117785 (NSF CAREER Award).

\bibliographystyle{plain} 
\bibliography{reference}

\end{document}